\documentclass[aps,prd,nofootinbib]{revtex4}
\usepackage{graphicx}
\usepackage{amsfonts}
\usepackage{amsmath}
\usepackage{amssymb}

\begin{document}

\title{Einstein Frame Regularization of the JNW Spacetime: From Brans-Dicke Singularity to Ellis Wormhole}
\author{A. Bhattacharya}
\email{bamrita323@gmail.com}
\affiliation{Department of Mathematics, Kidderpore College, 2, Pitamber Sircar Lane, Kolkata 700023, WB, India}
\author{R.N. Izmailov}
\email{izmailov.ramil@gmail.com}
\affiliation{Zel'dovich International Center for Astrophysics, Bashkir State Pedagogical University, 3A, October Revolution Street, Ufa 450077, RB, Russia}
\affiliation{Institute of Molecule and Crystal Physics, Ufa Federal Research Centre, Russian Academy of Sciences, Pr. Oktyabrya 151, Ufa, 450075, RB, Russia}
\author{R.Kh. Karimov}
\email{karimov_ramis_92@mail.ru}
\affiliation{Zel'dovich International Center for Astrophysics, Bashkir State Pedagogical University, 3A, October Revolution Street, Ufa 450077, RB, Russia}

\date{12 April 2025}

\begin{abstract}
We propose a novel way to achieve the regularization of the JNW globally naked singularity sourced by the canonical Einstein minimally coupled scalar field (EMS). The key idea is to complexify (Wick rotate) the JNW parameters, which yields a regular metric that surprisingly represents a twice asymptotically flat regular traversable wormhole, famously known as the Ellis class III wormhole sourced by the ghost EMS. We begin by arguing that the JNW solution belongs to a hierarchy of actions describing different gravity theories, which indicates that a regularized version of JNW automatically adds corresponding regular versions to the hierarchy of gravity theories.
\end{abstract}

%\pacs{}
\maketitle

%%%%%%%%%%%%%%%%%%  DATE  %%%%%%%%%%%%%%%%%%%

%%%%%%%%%%%%%%%%%%%%%%%%%%%%%%%%%%%%%%%%%
\section{Introduction}
\label{sec1}
%%%%%%%%%%%%%%%%%%%%%%%%%%%%%%%%%%%%%%%%%
The study of spacetime singularities has transcended its traditional role as a purely mathematical or foundational puzzle within general relativity to become a vibrant area of research with deep implications for modern cosmology \cite{Joshi:2002, Joshi:2011, Samanta:2018}. The discovery of the Universe's late-time acceleration \cite{Perlmutter:1999} has catalyzed an extensive exploration of cosmological models, many of which inevitably lead to singular behavior in both the past and the future \cite{Trivedi:2024}. This has established a direct conceptual link between the singularities formed in gravitational collapse, such as the Janis-Newman-Winicour (JNW) naked singularity, and the singularities predicted to occur in the evolutionary timeline of our Universe.

Recent advances have significantly broadened our understanding of singularities beyond the classic Big Bang. As extensively reviewed in \cite{Trivedi:2024}, a rich zoology of finite-time cosmological singularities has been classified. These singularities, which can occur in the context of dark energy models often driven by scalar fields \cite{Trivedi:2024, Trivedi:2022}, demonstrate that singular behavior is not an exotic exception but a generic feature of many viable cosmological scenarios. The very fact that a scalar field driving cosmic acceleration can lead to a future singularity in an FRW universe strengthens the motivation to fully understand scalar field configurations, like the JNW solution, that harbor naked singularities in a static, spherically symmetric setting.

Furthermore, naked singularities themselves are being actively investigated as potential astrophysical observables, offering an alternative to the black hole paradigm. A new class of static naked singularity solutions, derived from gravitational collapse, has been shown to admit a consistent two-fluid description and, crucially, to possess distinctive observational signatures in lensing and accretion disk properties that could, in principle, differentiate them from black holes \cite{Bhattacharya:2020}. This indicates that the study of naked singularities is not merely an academic exercise but is linked to the interpretation of current and future high-precision astrophysical observations.

The role of modified gravity and exotic matter configurations, which are central to our regularization procedure, also finds a natural home in cosmology. For instance, bouncing cosmologies, which aim to resolve the initial Big Bang singularity, often require a violation of the null energy condition (NEC) and exhibit singular behavior at the bounce point within the framework of modified gravity theories like $f(R,T)$ \cite{Sobhanbabu:2025}. This directly parallels our work, where achieving a regular, traversable wormhole from the singular JNW solution necessitates "exoticizing" the source matter (the ghost scalar field), which likewise violates the NEC. Finally, the foundational link between our solution-generating technique and alternative theories of gravity is echoed in cosmology. The Brans-Dicke (BD) theory, from which we trace the JNW solution's lineage, remains a cornerstone for studying varying gravitational constants and modified gravity effects in the early and late Universe \cite{Trivedi:2022}.

Regularizing a naked curvature singularity is always a challenge in general relativity. In an effort to meet this challenge, Pal et al. \cite{Pal:2023} in a recent article extended the Simpson and Visser (SV) "black-bounce" method \cite{Simpson:2019} of regularizing a black hole spacetime to cases where the initial metrics represent globally naked singularities such as the JNW metric \cite{Janis:1968, Virbhadra:1997} sourced by the Einstein minimally coupled scalar field and the Joshi--Malafarina--Narayan (JMN) metric \cite{Joshi:2011} representing a fluid model of naked singularity. In what follows, we are going to deal only with the JNW case. We regard the work in \cite{Pal:2023} as interesting and useful, except the shortcoming that the final spacetime is not fully regular but interpolates between a naked singularity and a wormhole depending on the associated parameters.

We wish to report here a novel alternative that does not have the above shortcoming yet converts the JNW naked singularity into a \textit{different} fully regularized final metric, which is an elegant, well-known, massive, twice asymptotically flat, traversable Ellis wormhole (more precisely, the Ellis class III wormhole \cite{Ellis:1973, Ellis:1974, Bronnikov:1973}\footnote{
Also called Ellis-Bronnikov wormhole because of their simultaneous but independent discovery of the metric.}). The wormhole spacetime preserves the original minimally coupled nature of the JNW source except adding before it an overall negative sign thereby "exoticizing" the source matter (ghost), as required by the wormhole topology. We shall take units such that $16\pi G=c=1 $.

%%%%%%%%%%%%%%%%%%%%%%%%%%%%%%%%%%%%%%%%%
\section{Naked singularity solutions}
\label{sec2}
%%%%%%%%%%%%%%%%%%%%%%%%%%%%%%%%%%%%%%%%%
It is important to note that the JNW solution belongs to a hierarchy of actions describing different gravity theories. Historically, Fisher \cite{Fisher:1948} in 1948 was the first to have derived the singular solution of the Einstein minimally coupled scalar (EMS) field theory in a certain gauge. Thereafter, other authors independently rediscovered its recovered form in different gauges. For instance, it is not often realized that the JNW solution is \textit{formally} equivalent to the vacuum BD class I solution \cite{Brans:1962} in the Jordan frame (JF), only re-expressed in the conformally rescaled Einstein frame (EF)\footnote{%
As an aside, we emphasize that the equivalence is strictly formal. The interpretations are different. In the JF, the scalar field $\varphi$ in Eq.(6) is a component of gravity in the vacuum BD theory, while in the EF, the scalar field $\phi $ in Eq.(10) is a matter field in Einstein's general relativity. However, the topic of the present report is different.}. It will be evident below that the EF incarnation of vacuum BD class I solution is variously known as the Buchdahl \cite{Buchdahl:1959} or JNW \cite{Janis:1968} or Ellis class I solution \cite{Ellis:1973, Ellis:1974, Bronnikov:1973}. These imply that a regularized version of JNW metric automatically adds corresponding regular versions of the solution to the repertoire of different gravity theories. Note that cosmological models in the BD theory do not show such naked singularities as we are about to regularize (see, e.g., \cite{Nimkar:2023}).

To gain a glimpse of such theories, we note that even the 4-dimensional, low energy effective action of heterotic string theory compactified on a 6-torus can be reduced to the BD action \cite{Kar:1999}. The tree level string action, keeping only linear terms in the constant string tension $\alpha ^{\prime }$ and in the Ricci curvature $\widetilde{\mathbf{R}}$, takes the following form in the matter free region ($S_{matter}=0$):%
\begin{equation}
S_{string}=\frac{1}{\alpha ^{\prime }}\int d^{4}x\sqrt{-\widetilde{g}}e^{-2%
\widetilde{\Phi }}\left[ \widetilde{\mathbf{R}}-4\widetilde{g}^{\mu \nu }%
\widetilde{\Phi }_{,\mu }\widetilde{\Phi }_{,\nu }\right] ,
\end{equation}%
where $\widetilde{g}^{\mu \nu }$ is the string metric and $\widetilde{\Phi}$ is the dilaton field. Note that the zero values of other matter fields do not impose any additional constraints either on the metric or on the dilaton \cite{Kar:1999}. Under the substitution
\footnote{%
The transformation $e^{-2\Phi} = \phi$ is a field redefinition that maps the BD scalar $\phi$ to the string-theory dilaton $\Phi$, revealing a close structural match between BD theory in the JF and the low-energy string effective action in the string frame \cite{Hrycyna:2013, Izmailov:2020}. This exponential parametrization is widely used because it naturally appears in string cosmology (e.g., pre–big bang scenarios \cite{Gasperini:2003, Conzinu:2025}), lets $\phi$ be interpreted as an effective gravitational coupling tied to the string coupling $g_{s} = e^{\langle\Phi\rangle}$, and makes frame dualities easier to analyze (expansion in one frame can look like contraction in another).}
\begin{equation}
e^{-2\widetilde{\Phi }}=\varphi ,
\end{equation}%
the above action reduces to the vacuum (matter-free) BD action%
\begin{equation}
S_{BD}=\int d^{4}x\sqrt{-\widetilde{g}}\left[ \varphi \widetilde{\mathbf{R}} - \frac{1}{\varphi }\widetilde{g}^{\mu \nu }\varphi _{,\mu }\varphi _{,\nu }\right] ,
\end{equation}%
in which the BD coupling parameter $\omega =-1$. This particular value is independent of specific models. It should however be noted that, Faraoni \cite{Faraoni:1998} discovered a very interesting property of the vacuum BD action, viz., it has a conformal invariance characterized by a constant gauge parameter $\xi $. Arbitrary values of $\xi $ can actually lead to any shift from the value $\omega =-1$. Thus, we keep $\omega $ arbitrary without loss of generality and consider the original matter-free BD action in the JF variables ($g_{\mu \nu },\varphi $) for constant $\omega$, which is (dropping tildes) 
\begin{equation}
S_{BD}=\int d^{4}x(-g)^{\frac{1}{2}}\left[ \varphi \mathbf{R}+\frac{\omega }{\varphi }g^{\mu \nu }\varphi _{,\mu }\varphi _{,\nu }\right] ,
\end{equation}%
where $\mathbf{R}$ is the Ricci scalar formed from $g_{\mu\nu}$. The BD field equations are%
\begin{equation}
\square^{2}\varphi =0,
\end{equation}
\begin{equation}
\mathbf{R}_{\mu\nu} - \frac{1}{2}g_{\mu\nu}\mathbf{R} = -\frac{\omega}{\varphi^{2}} \left[ \varphi _{,\mu} \varphi _{,\nu} - \frac{1}{2}g_{\mu\nu} \varphi _{,\sigma} \varphi ^{,\sigma}\right] - \frac{1}{\varphi} \left[\varphi _{;\mu} \varphi _{;\nu} - g_{\mu\nu} \square^{2} \varphi \right].
\end{equation}

To reach our goal in this paper, let us introduce what are called Dicke transformations, viz., 
\begin{eqnarray}
\overline{g}_{\mu \nu } &=&\varphi g_{\mu \nu } \\
d\phi &=&\sqrt{\frac{2\omega +3}{2\alpha }}\frac{d\varphi }{\varphi }; \quad \alpha \neq 0; \quad \omega \neq -\frac{3}{2},
\end{eqnarray}%
in which we have introduced, on purpose, a constant parameter $\alpha$ that can in principle have any sign. Then the action (4) takes the form of the EMS action in the EF variables ($\overline{g}_{\mu \nu },\phi $):%
\begin{equation}
S_{EMS}=\int d^{4}x\sqrt{-\overline{g}}\left[ \overline{\mathbf{R}}+\alpha 
\overline{g}^{\mu \nu }\phi _{,\mu }\phi _{,\nu }\right] .
\end{equation}%
If the EMS kinetic term in the action $\alpha \overline{g}^{\mu \nu }\phi_{,\mu }\phi _{,\nu }$ has an overall positive sign, the matter field is called canonical. If it has an overall negative sign, the matter field is called exotic or ghost, which can happen if $\alpha <0$ with $\phi $ real or $\alpha >0$ with $\phi $ imaginary. The EMS field equations are
\begin{eqnarray}
\overline{\mathbf{R}}_{\mu \nu } &=&-\alpha \phi _{,\mu }\phi _{,\nu } \\
\overline{\square }^{2}\phi &=&0.
\end{eqnarray}

The ansatz for BD solution to Eqs.(5,6), in isotropic coordinates ($t,r,\theta ,\varphi $) is taken as
\begin{equation}
ds^{2}=-e^{2\alpha (r)}dt^{2}+e^{2\beta (r)}dr^{2}+e^{2\nu (r)}r^{2}(d\theta ^{2}+\sin ^{2}\theta d\psi ^{2}).
\end{equation}%
The JF Brans class I solution \cite{Brans:1962} in the gauge $\beta -\nu =0$ is given by%
\begin{equation}
e^{\alpha (r)}=e^{\alpha _{0}}\left[ \frac{1-B/r}{1+B/r}\right] ^{\frac{1}{\lambda }},
\end{equation}%
\begin{equation}
e^{\beta (r)}=e^{\beta _{0}}\left[ 1+B/r\right] ^{2}\left[ \frac{1-B/r}{1+B/r}\right] ^{\frac{\lambda -C-1}{\lambda }},
\end{equation}%
\begin{equation}
\varphi (r)=\varphi _{0}\left[ \frac{1-B/r}{1+B/r}\right] ^{\frac{C}{\lambda}},
\end{equation}%
\begin{equation}
\lambda ^{2}\equiv (C+1)^{2}-C\left( 1-\frac{\omega C}{2}\right) >0,
\end{equation}%
where $\alpha _{0}$, $\beta _{0}$, $B$, $C$, and $\varphi _{0}$ are integration constants. The constants $\alpha _{0}$ and $\beta _{0}$ are determined by asymptotic flatness condition, viz., $\alpha _{0}=$ $\beta_{0}=0$. The solutions (13-16) can be rephrased, using the Dicke transformations (7) and (8), obtaining the EF version as%
\begin{eqnarray}
ds^{2} &=&-\left( 1+\frac{m}{2r}\right) ^{-2\beta }\left( 1-\frac{m}{2r}%
\right) ^{2\beta }dt^{2}+\left( 1-\frac{m}{2r}\right) ^{2(1-\beta )} \\
&&\times \left( 1+\frac{m}{2r}\right) ^{2(1+\beta )}[dr^{2}+r^{2}(d\theta
^{2}+\sin ^{2}\theta d\psi ^{2})]  \nonumber
\end{eqnarray}
\begin{equation}
\phi =\left[ \left( \frac{\omega +3/2}{\alpha }\right) \left( \frac{C^{2}}{%
\lambda ^{2}}\right) \right] ^{1/2}\ln \left[ \frac{1-\frac{m}{2r}}{1+\frac{m%
}{2r}}\right] ,
\end{equation}%
\begin{equation}
\beta =\frac{1}{\lambda }\left( 1+\frac{C}{2}\right) ,
\end{equation}%
where $m=2B$ is a redefined integration constant related to the asymptotic ADM mass $M$ by $M=m\beta $. The expression for $\lambda ^{2}$, of course, continues to be the same as Eq.(16), and using this, we can rewrite Eq.(18) as%
\begin{equation}
\phi =\left[ \frac{2(1-\beta ^{2})}{\alpha }\right] ^{1/2}\ln \left[ \frac{1-%
\frac{m}{2r}}{1+\frac{m}{2r}}\right] .
\end{equation}%
\textit{This is precisely the Ellis class I \cite{Ellis:1973, Ellis:1974, Bronnikov:1973} or Buchdahl \cite{Buchdahl:1959} solution} belonging to the field equations (10,11). Expanding the scalar field to first order, we have%
\begin{equation}
\phi \simeq \frac{q}{r},q=-m\left[ \frac{2(1-\beta ^{2})}{\alpha }\right]^{1/2}.
\end{equation}%
The stress-energy tensor for a massless minimally coupled scalar field $\phi $, viz.,%
\begin{equation}
T_{\mu \nu }=\alpha \left[ \phi _{,\mu }\phi _{,\nu }-\frac{1}{2}g^{\mu \nu
}\phi _{,\sigma }\phi ^{,\sigma }\right].
\end{equation}

In four-dimensional spacetime, the trace is
\begin{equation}
T = T^{\mu}_{ \mu} = g^{\mu\nu} T_{\mu\nu} = - \alpha \phi_{,\sigma} \phi^{,\sigma}.
\end{equation}
This makes the "strong-energy" combination especially simple:
\begin{equation}
T_{\mu\nu} - \frac{1}{2} T g_{\mu\nu} = \alpha \phi_{,\mu} \phi_{,\nu}.
\end{equation}
Below are the standard pointwise energy conditions implied by this stress-energy tensor. 

Null Energy Condition (NEC) \emph{definition}: For every null vector $k^{\mu}$ ($k^{\mu} k_{\mu} = 0$), $T_{\mu\nu}k^{\mu}k^{\nu} \geq 0$.

Using given $T_{\mu\nu}$ in (22), we can compute
\begin{equation}
T_{\mu\nu}k^{\mu}k^{\nu} = \alpha \left[\left(k^{\mu} \phi_{,\mu}\right)^{2} - \frac{1}{2} \left(k^{\mu} k_{\mu}\right) \left(\phi_{,\sigma} \phi^{,\sigma} \right) \right] = \alpha \left(k^{\mu} \phi_{,\mu}\right)^{2}.
\end{equation}
So, if $\alpha > 0$, the NEC satisfied (it can be saturated if $k\cdot \nabla \varphi = 0$).

Weak Energy Condition (WEC) \emph{definition}: For every timelike vector $u^{\mu}$, $T_{\mu\nu}u^{\mu}u^{\nu} \geq 0$.

In a local inertial frame (or by decomposing $\nabla \phi$ into parts parallel/orthogonal to $u^{\mu}$), one finds
\begin{equation}
T_{\mu\nu}u^{\mu}u^{\nu} = \frac{\alpha}{2} \left[\left(u^{\mu} \phi_{,\mu}\right)^{2} + \left(\nabla_{\perp}\phi\right)^{2} \right],
\end{equation}
where $\left(\nabla_{\perp}\phi\right)^{2} \geq 0$ is the squared magnitude of the part of $\nabla \phi$ orthogonal to $u$. Therefore, if $\alpha > 0$, the WEC satisfied for all timelike $u$.

Strong Energy Condition (SEC) \emph{definition}: For every timelike $u^{\mu}$, $\left(T_{\mu\nu} - \frac{1}{2}T g_{\mu\nu}\right) u^{\mu}u^{\nu} \geq 0$.

Using the identity above:
\begin{equation}
\left(T_{\mu\nu} - \frac{1}{2}T g_{\mu\nu}\right) u^{\mu}u^{\nu} = \alpha \left(\phi_{,\mu} u^{\mu} \right)^{2}.
\end{equation}
So, if $\alpha > 0$, the SEC satisfied (again, possibly saturated).

Dominant Energy Condition (DEC) \emph{definition}: For every future-directed timelike $u^{\mu}$, the energy-flux vector $-T^{\mu}_{ \mu} u^{\nu}$ is future-directed causal (non-spacelike). Equivalently (for type-I matter), in an orthonormal frame:
\begin{equation}
\rho \geq 0, \quad \rho \geq |p_{i}| \textmd{ for each principal pressure } p_{i}.
\end{equation}

For a massless scalar like this, one can always choose a local orthonormal frame aligned with $\nabla \phi$. In that frame the principal pressures satisfy $|p_{i}| = \rho$ (DEC is saturated) when $\alpha > 0$.

Next, for $\phi$ to be real, it is necessary
\begin{equation}
\frac{1-\beta^2}{\alpha} \geq 0.
\end{equation}
So, for a standard scalar field with $\alpha > 0$ (and hence NEC/WEC/SEC/DEC all satisfied), we get that
\begin{equation}
\beta^2 \leq 1.
\end{equation}

Thus, the stress-energy tensor (22) satisfies all energy conditions for $\alpha > 0$ and $\beta^2 \leq 1$. In the JF, this feature is not guaranteed, that's why the EF is often called "physical" for which the restriction $\omega >-3/2$ (which is the same as $\beta <1$) follows from Eq.(20) for $\phi $ to be real.

We wish to point out that the Ellis class I or Buchdahl solution shares many features of wormhole geometry. For instance, if we choose $\alpha <0$ and $\beta ^{2}>1$, this amounts to \textit{violating} all energy conditions by hand, while keeping $\phi $ real. Also, the Buchdahl metric (17) is invariant in form under inversion of the radial coordinate $r\rightarrow \frac{m^{2}}{4r}$ and so we have two asymptotically flat regions (at $r=0$ and $r=\infty $), on either side of the minimum area radius (throat) occurring at $r_{th}=\frac{m}{2}\left[ \beta +\sqrt{\beta ^{2}-1}\right]$. Alternatively, the same violations, hence wormhole geometry, can be achieved with $\alpha >0$, $\beta ^{2}>1$ in Eq.(20) or, equivalently, $\omega < -3/2, $ making $\phi $ imaginary. Thus, a real throat is guaranteed by the condition $\beta ^{2}>1$, which we may call the wormhole condition here. Also, one can verify that $\overline{\mathbf{R}}$ is negative for $\beta^{2}>1$ as expected of a wormhole geometry. However, despite all these facts, because of the occurrence of naked singularity at $r_{S}=m/2$, the wormhole is not traversable.

Using the coordinate transformation $l=r+\frac{m^{2}}{4r}$, the Ellis class I (or Buchdahl) solution (17) and (20) can be expressed as%
\begin{eqnarray}
ds^{2} &=&-f_{0}(l)dt^{2}+\frac{1}{f_{0}(l)}\left[ dl^{2}+(l^{2}-m^{2})%
\left( d\theta ^{2}+\sin ^{2}\theta d\psi ^{2}\right) \right] , \\
f_{0}(l) &=&\left( \frac{l-m}{l+m}\right) ^{\beta }, \\
\varphi _{0}(l) &=&\sqrt{\frac{\beta ^{2}-1}{2}}\ln \left[ \frac{l-m}{l+m}\right] .
\end{eqnarray}%
The Ellis class I solution in this form was discussed by Bronnikov and Shikin \cite{Bronnikov:1977} in which the naked singularity has now shifted to $l=m$. The throat $l_{th}$ appears at $l_{th}=r_{th}+\frac{m^{2}}{4r_{th}}=m\beta >m$ corresponding to $r=r_{th}$. Thus the minimum surface area has a value $4\pi m^{2}\beta ^{2}$.

The solutions (17) and (20) reduce, under the radial transformation $r=\rho
\left( {1+\frac{m}{{2\rho }}}\right) ^{2}$, to the JNW form considered by
Pal et al \cite{Pal:2023}: 
\begin{equation}
ds^{2}=-\left( {1-\frac{{2\eta }}{\rho }}\right) ^{m/\eta }dt^{2}+\left( 
{1-\frac{{2\eta }}{\rho }}\right) ^{-m/\eta }d\rho ^{2}+\left( {1-\frac{{%
2\eta }}{\rho }}\right) ^{1-m/\eta }\rho ^{2}d\Omega _{2}^{2},
\end{equation}%
\begin{eqnarray}
\phi (\rho ) &=&\sqrt{\frac{{1-\frac{{m^{2}}}{{\eta ^{2}}}}}{{2\alpha }}}\ln
\left( {1-\frac{{2\eta }}{\rho }}\right) 
\simeq \frac{q}{\rho }, \\
q &=&\left( \sqrt{\frac{{1-\beta }^{2}}{{2\alpha }}}\right) \times 2\eta,
\end{eqnarray}%
where we have put $m/\eta =\beta .$ For definiteness, we take $\alpha = 2$. Further identifications to reach the Virbhadra form \cite{Virbhadra:1997} are: $2\eta = b\Rightarrow b=\frac{2m}{\beta },$ and from $m/\eta =\beta $, one has for $\alpha = 2$, from the last Eq.(28), $\eta ^{2}-m^{2}=q^{2}$ so that $\beta = \frac{2m}{b}=\frac{2m}{2\sqrt{m^{2}+q^{2}}}$.

%%%%%%%%%%%%%%%%%%%%%%%%%%%%%%%%%%%%%%%%%
\section{Regularizing via Wick rotation}
\label{sec3}
%%%%%%%%%%%%%%%%%%%%%%%%%%%%%%%%%%%%%%%%%
From the foregoing, it is quite evident that the metrics (13-16), (17,20), (31-33) possess in different theories the same naked singularity as appears in the more familiar JNW form (34-36). To regularize it, for convenience we consider the Ellis class I solutions ($f_{0}$,$\varphi _{0}$) in Eq.(32,33) and analytically continue it while maintaining the real, positive numerical value of the throat radius $l_{th}=m\beta $. In the solution set ($f_{0}$,$\varphi _{0}$), we thus Wick rotate the parameters as 
\begin{equation}
m\rightarrow -im, \quad \beta \rightarrow i\beta
\end{equation}%
so that $l_{th}=m\beta $ is invariant in sign and magnitude. Then the metric obtained from the seed solution ($f_{0}$,$\varphi _{0}$) becomes%
\begin{equation}
ds^{2}=-f_{0}^{\prime }(l)dt^{2}+\frac{1}{f_{0}^{\prime }(l)}\left[
dl^{2}+(l^{2}+m^{2})\left( d\theta ^{2}+\sin ^{2}\theta d\psi ^{2}\right) %
\right],
\end{equation}%
\begin{equation}
f_{0}^{\prime }(l) = \exp \left[ -2\beta \textmd{arccot} \left(\frac{l}{m}\right) \right],
\end{equation}%
\begin{equation}
\varphi _{0}^{\prime }(l)=\left[ \sqrt{1+\beta ^{2}}\right] \textmd{arccot}%
\left( \frac{l}{m}\right).
\end{equation}%
\textit{This is just the regular Ellis class III wormhole solution in \cite{Ellis:1973, Ellis:1974, Bronnikov:1973}}, which can be rephrased in its original form by using the identities:%
\begin{eqnarray}
\textmd{arccot}(x)+\arctan (x) &=&+\frac{\pi }{2};x>0 \\
&=&-\frac{\pi }{2};x<0
\end{eqnarray}%
and the function on the left shows a finite jump (of magnitude $\pi $) at $x=0$. Thus, we get from Eqs.(39) and (40) two branches, the $+ve$ sign corresponds to the side $l>0$ and the $-ve$ sign to $l<0$:
\begin{equation}
f_{0\pm }^{Ellis}(l)=\exp [-2\beta \{\pm \frac{\pi }{2}-\arctan (\frac{l}{m})\}],
\end{equation}%
\begin{equation}
\varphi _{0\pm }^{Ellis}(l)=\left[ \sqrt{1+\beta ^{2}}\right] \left( \pm \frac{\pi }{2}-\arctan (\frac{l}{m})\right).
\end{equation}%
We might study the solutions (39) and (40) \textit{per se} that has no branches but has a finite jump at the origin because $f_{0}^{\prime}(l)\rightarrow e^{\pm \pi \beta }$ as $l\rightarrow \pm 0$, while there is no asymptotic mass jump since $f_{0}^{\prime }(l)\rightarrow 1$ as $l\rightarrow \pm \infty $. We disregard this form as the finite jump at the origin could prevent traversability.

On the other hand, notice that each of the $\pm $ branches in Eqs.(43), (44) exactly satisfies the EMS equations for unrestricted range of $l$ with no discontinuity at $l=0$. \textit{In fact, what is known in the literature as the Ellis class III solution \cite{Ellis:1973, Ellis:1974, Bronnikov:1973} is just the }$+ve$\textit{\ branch, viz., }$\left( f_{0+}^{Ellis},\varphi _{0+}^{Ellis}\right) $\textit{,} which is continuous over the entire interval $l\in (-\infty ,+\infty )$. Each of the individual branch represents a geodesically complete, asymptotically flat traversable wormhole (termed "drainhole" by Ellis) having different masses, one positive and the other negative, on two sides of the throat respectively. The $-ve$ branch is also equally good possessing exactly the same properties as the $+ve$ branch. As a byproduct, we see that the Ellis class I and III solutions are strictly \textit{not} independent solutions of the EMS theory as one can be obtained from the other by Wick rotation, however with a great difference that one is a naked singularity and the other is a regular traversable wormhole.

The Ellis class III metric function $f_{0+}^{Ellis}(l)\rightarrow 1$ as $l\rightarrow +\infty $ but $f_{0+}^{Ellis}(l)\rightarrow e^{-2\pi \beta }$ as $l\rightarrow -\infty $. These two limits correspond to a Schwarzschild mass $M$ at one mouth and $-Me^{\pi \beta }$ at the other. There is no discontinuity at the origin because $f_{0+}^{Ellis}(l)\rightarrow e^{-\pi \beta }$ as $l\rightarrow \pm 0$. The curvature scalar is given by%
\begin{equation}
\mathbf{R}_{0+}^{Ellis}=-\frac{2m^{2}(1+\beta ^{2})}{(l^{2}+m^{2})^{2}}\exp \left[ -2\beta \{\frac{\pi }{2}-\arctan \left( \frac{l}{m}\right) \}\right],
\end{equation}%
which goes to zero as $l\rightarrow \pm \infty $. That means that the spacetime is flat on two sides of the Ellis class III wormhole. Next, the singular coordinate radius ($r_{S}=m/2$) has now shifted to the origin $l=r - \frac{m^{2}}{4r}=0$, where the curvature is finite meaning that the singularity is regularized. Also, there is no jump at the origin, $R_{0+}^{Ellis}\rightarrow -\frac{2(1+\beta ^{2})}{m^{2}}e^{-\pi \beta}$ as $l\rightarrow \pm 0$. The area radius too does not show such jump either, $\rho _{0+}^{Ellis}(l)=\sqrt{f_{0+}^{-1(Ellis)}(l^{2}+m^{2})}\rightarrow m\sqrt{e^{\pi \beta }}$ as $l\rightarrow \pm 0$. These show that Ellis class III wormhole $\left( f_{0+}^{Ellis},\varphi _{0+}^{Ellis}\right) $ is traversable and has all the desirable properties of a regular wormhole. Similar considerations apply for the $-ve$ branch.

%%%%%%%%%%%%%%%%%%%%%%%%%%%%%%%%%%%%%%%%%
\section{Conclusion}
\label{sec4}
%%%%%%%%%%%%%%%%%%%%%%%%%%%%%%%%%%%%%%%%%
To conclude, we illustrated how the canonical EMS, and hence its solution JNW naked singularity, is linked to actions of different gravity theories. Thus, regularizing the JNW singularity in the EF means that the same regularization can very well apply to progenitor theories such as the JF BD theory or string theory. An inevitable feature of the regularization procedure is the emergence of ghost matter threading the final spacetime. This fact is evident from the works in \cite{Pal:2023, Bronnikov:2022} that applied the SV procedure to the initial JNW metric obtaining a final metric sourced, either partially or fully, by ghost matter. The present procedure uses the complex Wick rotation of JNW parameters $m$ and $\beta$ of (31-33), which magically converts the naked singularity into a nice object - the famous Ellis class III regular twice asymptotically flat traversable wormhole sourced by a ghost EMS. Thus the method provides an elegant way for regularizing the JNW singularity free of the shortcoming alluded to earlier, that is, the final wormhole (43, 44) is regular for \textit{all} values of the seed JNW parameters.

\end{document}